\documentclass[12pt]{article}

\usepackage{array} 
\usepackage{amssymb}
\usepackage{graphics,graphpap}
\usepackage{graphicx}
\usepackage{color}
\usepackage{graphicx}
\usepackage{dcolumn}
\usepackage{epsfig}
\usepackage{epstopdf}
\usepackage{bbm}
\usepackage{amsmath}
\usepackage{amsfonts}
\usepackage{textcomp}
\usepackage{setspace}
\usepackage{slashed}

\usepackage{todonotes}

\usepackage{cite}

\usepackage{url}

\usepackage{booktabs}

\newcommand{\beq}{\begin{eqnarray}}
\newcommand{\eeq}{\end{eqnarray}}

\newcommand{\bmp}{\noindent\begin{minipage}{16cm}}
\newcommand{\emp}{\end{minipage}\vskip 7mm} 

\def\drawbox#1#2{\hrule height#2pt
        \hbox{\vrule width#2pt height#1pt \kern#1pt
              \vrule width#2pt}
              \hrule height#2pt}

\def\Asym#1#2{\vcenter{\vbox{\drawbox{#1}{#2}
              \kern-#2pt 
              \drawbox{#1}{#2}}}}

\def\simge{\mathrel{%
   \rlap{\raise 0.511ex \hbox{$>$}}{\lower 0.511ex \hbox{$\sim$}}}}

\def\simle{\mathrel{
   \rlap{\raise 0.511ex \hbox{$<$}}{\lower 0.511ex \hbox{$\sim$}}}}

\def\s#1{\setbox0=\hbox{$#1$}%
\rlap{\ifdim\wd0>.7em\kern.22\wd0\else\kern.1\wd0\fi /}#1}

\begin{document}

\begin{titlepage}
\title{\vspace*{-2.0cm}
\hfill {\small MPP-2015-20}\\[20mm]
\bf\Large
Parity Problem of the Scotogenic Neutrino Model
\\[5mm]\ }

\author{
Alexander Merle\thanks{email: \tt amerle@mpp.mpg.de}~~~and~~Moritz Platscher\thanks{email: \tt mplat@mpp.mpg.de}
\\ \\
{\normalsize \it Max-Planck-Institut f\"ur Physik (Werner-Heisenberg-Institut),}\\
{\normalsize \it F\"ohringer Ring 6, 80805 M\"unchen, Germany}\\
}
\date{\today}
\maketitle
\thispagestyle{empty}

\begin{abstract}
\noindent
We show that Ma's scotogenic model, which is arguably one of the simplest settings containing a Dark Matter candidate and generating a naturally suppressed active neutrino mass at 1-loop level, suffers from a potentially severe hierarchy-type problem. In case the right-handed neutrinos involved have sufficiently large masses, they can via loop effects drive the mass parameter of the inert scalar contained in the model towards negative values. This behaviour leads to a breaking of the $\mathbb{Z}_2$ parity symmetry built into the model which is paramount to keeping the setting consistent at low energies -- without it  the model would lose its Dark Matter candidate and the neutrino mass would not be naturally suppressed. Thus, if the breaking occurs at a sufficiently low scale, it could potentially spoil the success of the whole model. Trying to avoid this consistency problem leads to a new constraint on the model parameter space which has not yet been described in the literature.
\end{abstract}

\end{titlepage}

\section{\label{sec:into}Introduction}

The smallness of neutrino masses and the identity of the Dark Matter (DM) are among the greatest puzzles of modern particle physics, as is the question of how to extend our Standard Model (SM) and how to probe the extensions. The most attractive theories beyond the SM are those which can address several of the known open problems. This is often the case for models generating a light neutrino mass only at loop level (i.e., radiatively) -- cf.\ e.g.\ Ref.~\cite{King:2014uha} to see how such settings are constrained at both, low and high energies. Depending on the particle content, there exist models which generate an active neutrino mass at 1-loop~\cite{Zee:1980ai}, 2-loop~\cite{Zee:1985id,Babu:1988ki}, or 3-loop~\cite{Gustafsson:2012vj,Gustafsson:2014vpa} level, but probably the simplest extension compatible with all data is Ma's scotogenic model~\cite{Ma:2006km}.
 
The scotogenic model just adds three right-handed (RH) neutrinos $N_R$ and a scalar doublet $\eta$ to the SM, all of which are charged under an additional $\mathbb{Z}_2$ parity symmetry.\footnote{Due to the $\mathbb{Z}_2$ symmetry restricting its interactions, the new scalar doublet is usually called \emph{inert}.} This symmetry is crucial for the model to work; without it, neutrino masses would already be generated at tree level and none of the possible DM candidates of the model could be stable. However, if the $\mathbb{Z}_2$ is intact, the scotogenic model cannot only account for phenomenologically valid neutrino masses~\cite{Sierra:2008wj,Suematsu:2009ww,Ahn:2012cga} and potentially for DM~\cite{LopezHonorez:2006gr,Sierra:2008wj,Suematsu:2009ww,Dolle:2009fn,Gelmini:2009xd}, it can also lead to a variety of phenomena in low-energy experiments such as lepton flavour and/or number violation~\cite{Kubo:2006yx,Sierra:2008wj,Suematsu:2009ww,Adulpravitchai:2009gi,Toma:2013zsa,Vicente:2014wga} or in high-energy 
collider searches~\cite{Cao:2007rm,BarShalom:2008gt,Dolle:2009ft,Arhrib:2012ia}, as well as to new aspects for neutrino model building~\cite{Ma:2008ym,Adulpravitchai:2009re,Ahn:2012cga}.

A few years ago, the first study of the renormalisation group running of the scotogenic model has appeared in Ref.~\cite{Bouchand:2012}. It has been shown that several parameters exhibit strong running effects. This is known in similar settings: Ref.~\cite{Hambye:2007vf} showed that an inert scalar can trigger electroweak symmetry breaking (EWSB) at 1-loop level even if not present at tree level -- a fact that remains true if the setting is extended~\cite{Kadastik:2009ca} or discretised~\cite{Lewis:2010ps}. More generally it is well-known  that scalar mass parameters are sensitive to large scales~\cite{Malinsky:2012tp}, which is also true for models with two Higgs doublets~\cite{Biswas:2014uba}.

In this paper we apply a similar logic, i.e., we investigate corrections to the mass parameter of the inert scalar. The heavy RH neutrinos can drive the squared mass parameter of the inert scalar towards negative values, thereby giving a non-zero vacuum expectation value (VEV) $\left\langle\eta\right\rangle$ and by that \emph{breaking the crucial $\mathbb{Z}_2$ parity symmetry}, which is why we call this observation the \emph{parity problem} of the scotogenic model. This could be disastrous given that, if the parity was not a conserved global quantum number, not only would the model lose the stability of its DM candidates but also a neutrino mass would be generated at tree level at phenomenologically unacceptably large values.
 
On the other hand, given that the breaking happens at a high scale, the question is justified why we should bother for low-energy phenomenology. The answer is two-fold: first, broken discrete symmetries could potentially lead to problems in cosmology. If the $\mathbb{Z}_2$ symmetry was e.g.\ intact at a high scale but then at some point broken by the large RH neutrino masses, we would potentially be in danger of creating unwanted domain walls~\cite{Zeldovich:1974uw}, which could modify the history of the Universe in an undesirable way. At the same time, a VEV of the inert scalar would lead to large tree level masses for active neutrinos. This would cause them to immediately freeze-out and possibly lead to an intermediate matter-dominated phase of the Universe, which would again alter the expansion history. However, one could argue that the domain wall problem can be cured~\cite{Preskill:1991kd,Riva:2010jm,Dvali:1995cc,Dvali:1996zr,Larsson:1996sp} and that active neutrinos may simply re-thermalise as soon as 
the $\mathbb{Z}_2$ is intact again. In addition, the symmetry will be restored at sufficiently high temperatures~\cite{Quiros:1999jp}. Thus, given that all these processes happen before big bang nucleosynthesis, there may not even be an observable remnant. This is a perfectly justified viewpoint, but even if one disregards the points above it turns out that the breaking scale of the $\mathbb{Z}_2$ can be as low as a few TeV, as we will show in Sec.~\ref{sec:Numerics_BScale}. If that is the case, DM production can be significantly modified in the scotogenic model and points in the parameter space which are thought to be consistent and to lead to successful DM production could in fact be in trouble due to the parity problem. Ultimately, the message is that one has to be careful and check for any given parameter point whether a potential $\mathbb{Z}_2$ breaking leads to a problem, or not.

As a side note, it is interesting to mention that obviously any issues associated with a too large RH neutrino mass scale could be avoided if all RH neutrino masses were sufficiently small. From this point of view, one could even argue that to some extend the scotogenic model would have a preference for light sterile neutrino masses. Since they would then also be lighter than all neutral scalar components, they would automatically be the actual DM candidates of the model. Such settings are known to work very well (see e.g.\ Refs.~\cite{Dodelson:1993je,Shi:1998km,Bezrukov:2009th,Nemevsek:2012cd,Kusenko:2006rh,Petraki:2007gq,Merle:2013wta,Adulpravitchai:2014xna,Merle:2015oja,Frigerio:2014ifa,Lello:2014yha,Abada:2014zra,Boyanovsky:2008nc,Shuve:2014doa} for suitable production mechanisms), and they have also been discussed from a phenomenological point of view in the context of the scotogenic model~\cite{Sierra:2008wj,Gelmini:2009xd}. While it is not easy to find a suitable mechanism to motivate light sterile 
neutrino masses in the scotogenic model~\cite{Merle:2013gea}, the above arguments could be interpreted as such a scenario with light sterile neutrinos being in fact quite natural.

This paper is organised as follows. After a brief overview of the model in Sec.~\ref{sec:Overview}, we discuss in Sec.~\ref{sec:Vacuum} the general possibilities for its possible vacuum configurations and illustrate the approximate constraints arising from avoiding a violation of the $\mathbb{Z}_2$ parity in the scotogenic model, which yields a simple but accurate formula. A more advanced numerical analysis of a few detailed examples is presented in Sec.~\ref{sec:Numerics}. We discuss some of the aforementioned implications for cosmology in Sec.~\ref{sec:Discussion}, before we conclude in Sec.~\ref{sec:Summary}. Technical details can be found in the Appendix.

\section{\label{sec:Overview}Model Overview \& Constraints}

The \emph{scotogenic model}~\cite{Ma:2006km} is one of the most minimal frameworks combining a naturally small neutrino mass at 1-loop level with several DM candidate particles. The particle content is that of the SM, supplemented by (typically) three RH neutrinos $N^i_R$ ($i=1,2,3$) and a second scalar doublet $\eta$ with SM quantum numbers identical to those of the Higgs. The crucial addendum is an additional $\mathbb{Z}_2$ (parity) symmetry, under which all SM particles are neutral whereas the new fields are odd. It is this symmetry which simultaneously leads to the light neutrino mass being generated at 1-loop level only and to the stability of the potential DM candidates.

Compared to the SM, several qualitatively new terms appear in the Lagrangian:
\begin{itemize}

\item The RH neutrinos get a direct Majorana mass term $\frac{1}{2} \overline{N_R^i} M_{ij} {N_R^j}^c + h.c.$, which leads to masses $M_{1,2,3}$ upon diagonalisation.

\item A neutrino Yukawa coupling $ \mathcal{L}_{\rm Yukawa} \supset - h_{ij} \overline{N^i_R} \tilde{\eta}^\dag {\ell^j_L} + h.c. \ (\tilde{\eta}=i \sigma_2 \eta^*)$ involving the new scalar and the RH neutrinos in addition to the SM lepton doublets $\ell^j_L$. It is crucial to observe that this term does \emph{not} lead to a tree level neutrino mass, as long as the $\mathbb{Z}_2$ is unbroken and thus prevents the field $\eta$ from obtaining a VEV.

\item The scalar potential involving the SM Higgs $\phi$ and the $\eta$ field is given by:
\begin{equation} \label{eq:potential}
\begin{aligned}
  V &= m_1^2 \phi^\dag\phi + m_2^2 \eta^\dag\eta + \frac{\lambda_1}{2} \left(\phi^\dag\phi\right)^2 + \frac{\lambda_2}{2} \left(\eta^\dag\eta\right)^2 \\
    & + \lambda_3 \left(\phi^\dag\phi\right)\left(\eta^\dag\eta\right) + \lambda_4\left(\eta^\dag\phi\right)\left(\phi^\dag\eta\right) + \frac{\lambda_5}{2} \left[(\eta^\dag\phi)^2 + h.c.\right].
\end{aligned}
\end{equation}
In this expression both mass parameters $m_{1,2}^2$ must be real, as need to be the couplings $\lambda_{1,2,3,4}$; $\lambda_5$ can be chosen real and positive by absorbing its phase into $\eta$.

\end{itemize}
Note that it is the combination of the Majorana mass term, the new Yukawa coupling, and the $\lambda_5$-term in Eq.~\eqref{eq:potential} which violates lepton number. If any of those coefficients was zero, a global $U(1)$ lepton number could be consistently defined and the symmetry of the Lagrangian would be increased. Thus, by virtue of 't Hooft naturalness~\cite{'tHooft:1979bh}, the renormalisation group equations (RGEs) for those quantities will only allow for changes proportional to the couplings themselves, so that they remain small everywhere if they are small at any energy scale, cf.\ the Appendix.

The scalar potential~\eqref{eq:potential} needs to yield EWSB without compromising the $\mathbb{Z}_2$ parity. This suggests the parametrisation $\phi = \left(0,v + \frac{h}{\sqrt{2}}\right)^T$ and $\eta=\left(\eta^+, \eta^0\right)^T$, where only the physical fields are written down explicitly. Splitting $\eta^0$ into real and imaginary parts, $\eta^0 = \frac{1}{\sqrt{2}}\left(\eta_R+i\eta_I\right)$, gives rise to the following physical scalar masses~\cite{Dolle:2009}:
\begin{subequations}
\begin{align} 
	m_h^2 &= 2 \lambda_1 v^2 \label{eq:masses1},\\
	m_\pm^2 &= m_2^2 + v^2 \lambda_3 \label{eq:masses2}, \\
	m_R^2 &= m_2^2 + v^2 \left(\lambda_3 + \lambda_4 + \lambda_5 \right) \label{eq:masses3}, \\
	m_I^2 &= m_2^2 + v^2 \left(\lambda_3 + \lambda_4 - \lambda_5 \right) \label{eq:masses4}.
\end{align}
\end{subequations}
Note that the parameters in the scalar potential are subject to a number of theoretical constraints originating from the requirement that the scalar potential be bounded from below~\cite{Branco:2011,Maniatis:2006,Klimenko:1984}:
\begin{equation} \label{eq:BoundedFromBelow}
 \lambda_1 > 0\, , \quad \lambda_2 > 0\, , \quad \lambda_3 > - \sqrt{\lambda_1 \lambda_2}\,,\quad \lambda_3 + \lambda_4 - |\lambda_5|>- \sqrt{\lambda_1 \lambda_2},
\end{equation}
which also affects the valid mass spectra resulting from Eqs.~\eqref{eq:masses1} to~\eqref{eq:masses4}. In addition, we can impose the condition that we remain in a perturbative regime, that is $\lambda_{1,2,3,4,5} \lesssim \mathcal{O}(1)$.\footnote{Note that there are some ambiguities in the definition of perturbativity~\cite{Nebot:2007bc,Herrero-Garcia:2014hfa}. For definiteness, we have chosen the criterion $\lambda_{1,2,3,4,5} < 4\pi$ in our numerical computations.} In our numerical analysis, we demand that the above conditions remain valid for the 1-loop corrected running couplings up to a given high energy scale. The 1-loop RGEs can be found in the Appendix.

\section{\label{sec:Vacuum}Vacuum Structure \& New Constraints}

The scalar sector of the scotogenic model is a particular realisation of a two Higgs doublet model (THDM) with a $\mathbb{Z}_2$ symmetry imposed on some fields. In a general THDM both scalar doublets (called $\phi_1$ and $\phi_2$ here to ease the comparison) may acquire a VEV and we find the tree level vacuum of the theory by minimising Eq.~\eqref{eq:potential}. Replacing the fields by their VEVs $v_{1,2}$, which can be chosen real if both electric charge and CP are conserved~\cite{Branco:2011,peskin1995}, leads to two equations:
\begin{equation}
v_1 \left(m_1^2+\lambda_1 v_1^2 + \lambda v_2^2 \right) = 0\ \ \ {\rm and}\ \ \ v_2 \left(m_2^2+\lambda_2 v_2^2 + \lambda v_1^2 \right) = 0,
\end{equation}
where we have abbreviated $\lambda \equiv \lambda_3 +\lambda_4 + \lambda_5$.

These equations allow four qualitatively different sets of VEVs:\footnote{\textcircled{\raisebox{-.2pt}{\scriptsize2}} and \textcircled{\raisebox{-.2pt}{\scriptsize3}} each have two real solutions, which are physically equivalent due to global symmetries, while \textcircled{\raisebox{-.2pt}{\scriptsize4}} gives rise to four solutions, which reduce to two physically \emph{inequivalent} solutions~\cite{Barroso:2013awa}.}
\begin{enumerate}
	\item[\textcircled{\raisebox{-.2pt}{\scriptsize1}}] $v_1^2 = v_2^2 = 0$,
	\item[\textcircled{\raisebox{-.2pt}{\scriptsize2}}] $v_1^2 = - \frac{m_1^2}{\lambda_1}, \quad v_2^2=0$,
	\item[\textcircled{\raisebox{-.2pt}{\scriptsize3}}] $v_1^2=0, \quad v_2^2 = - \frac{m_2^2}{\lambda_2}$,
	\item[\textcircled{\raisebox{-.2pt}{\scriptsize4}}] $v_1^2 = \frac{\lambda_2 m_1^2 - \lambda m_2^2}{\lambda^2 - \lambda_1 \lambda_2}$, \quad $v_2^2 = \frac{\lambda_1 m_2^2 - \lambda m_1^2}{\lambda^2 - \lambda_1 \lambda_2}$.
\end{enumerate}
While \textcircled{\raisebox{-.2pt}{\scriptsize1}} and \textcircled{\raisebox{-.2pt}{\scriptsize2}} respect the $\mathbb{Z}_2$, \textcircled{\raisebox{-.2pt}{\scriptsize3}} and \textcircled{\raisebox{-.2pt}{\scriptsize4}} break it spontaneously. In our numerical analysis, we have investigated whether \textcircled{\raisebox{-.2pt}{\scriptsize1}} or \textcircled{\raisebox{-.2pt}{\scriptsize2}} are stable minima of the potential. Otherwise the $\mathbb{Z}_2$ would be broken in any case since the potential must have a minimum and only $\mathbb{Z}_2$ breaking vacua are left.

Computing the Hessians we find that we have a stable, $\mathbb{Z}_2$ symmetric vacuum if
\begin{equation} \label{eq:Z2broken}
  m_2^2 \ge 
    \begin{cases}
      0 & \text{if } m_1^2 \ge 0 \Leftrightarrow \textcircled{\raisebox{-.2pt}{\scriptsize1}},\\
      \frac{\lambda}{\lambda_1} m_1^2 & \text{if } m_1^2 < 0 \Leftrightarrow \textcircled{\raisebox{-.2pt}{\scriptsize2}}.
    \end{cases}
\end{equation}
The latter equation is equivalent to the condition $m_R^2 \ge 0$, i.e.\ the field $\eta_R$ develops a VEV if its mass square -- an eigenvalue of the Hessian -- becomes negative. Had we allowed for a relative phase between $v_1$ and $v_2$, we would find that ${\rm Im}~\phi_2$ (i.e.\ $\eta_I$) may develop a VEV. The condition to exclude this is $m_I^2 \ge 0$. Similarly, to avoid breaking electric charge, we need $m_\pm^2 \ge 0$. If we ignore a possible instability of the $\mathbb{Z}_2$ symmetric vacua, we may expand the theory around the wrong vacuum. Expanding around the correct vacuum, however, would alter the phenomenological predictions of the model and compromise the $\mathbb{Z}_2$ symmetry.

Avoiding vacuum instability, of course, does not exclude that the $\mathbb{Z}_2$ symmetric minima are not the \textit{global} minimum of the potential. We can investigate this by plugging the solutions \textcircled{\raisebox{-.2pt}{\scriptsize1}} -- \textcircled{\raisebox{-.2pt}{\scriptsize4}} back into the scalar potential, Eq.~\eqref{eq:potential}, which yields~\cite{Ginzburg:2010wa}:
\begin{subequations} \allowdisplaybreaks
\begin{gather}
  V_\textrm{\textcircled{\raisebox{-.7pt}{\scriptsize1}}} \equiv V\left(v_1^2=0,v_2^2=0\right) = 0,\label{eq:GlobalMin1}\\
  V_\textrm{\textcircled{\raisebox{-.7pt}{\scriptsize2}}} \equiv V\left(v_1^2=-\frac{m_1^2}{\lambda_1},v_2^2=0\right) = - \frac{m_1^4}{2\lambda_1},\label{eq:GlobalMin2}\\
  V_\textrm{\textcircled{\raisebox{-.7pt}{\scriptsize3}}} \equiv V\left(v_1^2=0,v_2^2=-\frac{m_2^2}{\lambda_2}\right) = - \frac{m_2^4}{2\lambda_2},\label{eq:GlobalMin3}\\
  V_\textrm{\textcircled{\raisebox{-.7pt}{\scriptsize4}}} \equiv V\left(v_1^2 = \frac{\lambda_2 m_1^2 - \lambda m_2^2}{\lambda^2 - \lambda_1 \lambda_2},v_2^2 = \frac{\lambda_1 m_2^2 - \lambda m_1^2}{\lambda^2 - \lambda_1 \lambda_2}\right) = \frac{\lambda_2 m_1^4 - 2\lambda m_1^2 m_2^2 + \lambda_1 m_2^4}{2(\lambda^2-\lambda_1\lambda_2)}.\label{eq:GlobalMin4}
\end{gather}
\end{subequations}
Any of the above solutions \textcircled{\raisebox{-.2pt}{\scriptsize1}} -- \textcircled{\raisebox{-.2pt}{\scriptsize4}} is the global minimum if $v_{1,2}^2 \ge 0$ and if it has the lowest vacuum energy, $V_\textrm{\textcircled{\raisebox{-.7pt}{\scriptsize $i$}}}<V_\textrm{\textcircled{\raisebox{-.7pt}{\scriptsize $k$}}}$ for all $k\neq i$. Note that, however, there is some freedom in choosing a minimum which is not the global one if the decay time is larger than the age of the Universe, see e.g.~\cite{Barroso:2013awa}. Nevertheless, we can use the above equations to check whether the breaking of $\mathbb{Z}_2$ is the only constraint, or if there could be further constraints, which originate from the minima \textcircled{\raisebox{-.2pt}{\scriptsize1}} and \textcircled{\raisebox{-.2pt}{\scriptsize2}} not being the global one.\\

For illustrative purposes, we now show analytically that the $\mathbb{Z}_2$ symmetry can be broken spontaneously due to radiative corrections. We do so by directly calculating the breaking scale. To simplify the equations, we only consider one generation of fermions and assume $m_1^2(\mu)>0$ for all scales $\mu$, i.e.\ we study under which conditions only $m_2^2(\mu)$ becomes negative. For sufficiently small quartic scalar couplings we can ignore their contributions to the RGEs altogether and consider two simple but illustrative limiting cases, where all quantities under consideration are assumed to be real. In addition, we shall assume that the Majorana mass $M$ does not run at all, which is justified numerically, cf.\ Ref.~\cite{Bouchand:2012}.
 
\paragraph{Case 1: Large neutrino Yukawa coupling -- $h(\mu) \gg g_i(\mu)$}
${}$\\
In this limiting case, neglecting any gauge and non-neutrino Yukawa couplings, we can approximate the coupled RGEs~\eqref{eq:neutrinoRG} and~\eqref{eq:m2RG} for $h$ and $m_2^2$, respectively, as follows:
\begin{equation}
 \mathcal{D}h \simeq \frac{5}{2} h^3\,, \qquad \mathcal{D}m_2^2 \simeq 2 h^2 m_2^2 - 4 h^2 M^2,
 \label{eq:RGE1}
\end{equation}
where $\mathcal{D} \equiv 16 \pi^2 \mu \frac{\textrm{d}}{\textrm{d}\mu} \equiv 16 \pi^2 \frac{\textrm{d}}{\textrm{d}t}$ and we have suppressed explicit scale dependence for brevity. These differential equations can be solved exactly and yield a condition for the mass square to become negative, i.e.\ for symmetry breaking to occur at some point $t=t_*$,
\begin{equation}
 m_2^2(t_*) \overset{!}{=} 0 \quad \Leftrightarrow \quad \left(\frac{5}{16 \pi^2} h^2(0)\right)t_* = 1-\left(1-\frac{m_2^2(0)}{2M^2}\right)^{5/2}.
\end{equation}
This yields in the linear approximation [where $m_2^2(0) \ll 2 M^2$]:
\begin{equation} \label{eq:condition1}
 t_* \simeq \frac{4\pi^2 m_2^2(0)}{M^2 h^2(0)}.
\end{equation}

\paragraph{Case 2: Small neutrino Yukawa coupling -- $h(\mu) \ll g_i(\mu)$}
${}$\\
In this case the RGEs involve the gauge couplings Eq.~\eqref{eq:gauge-RGE}:
\begin{equation}
 \mathcal{D} h \simeq -\frac{3}{4}\left(g_1^2 + 3 g_2^2\right) h,\ \ \ \mathcal{D} m_2^2 \simeq -\frac{3}{2}\left(g_1^2 + 3 g_2^2\right)m_2^2 - 4 h^2 M^2.
 \label{eq:RGEh2}
\end{equation}
The solutions are simple to find and we get
\begin{equation} \label{eq:condition2}
 t_* = \frac{4 \pi^2 m_2^2(0)}{M^2 h^2(0)}.
\end{equation}
Note that the conditions for both cases agree if $m_2^2(0) \ll 2 M^2$, which is just the case we are interested in. 

One might object that a $\mathbb{Z}_2$ broken above some large scale $t_*$ is irrelevant as long as the low energy phenomenology is unaffected, in which case we should only be concerned whether $t_*$ is large enough for this to be true. On the other hand, one may argue that a broken symmetry in the UV is undesirable from an aesthetical point of view, and we can impose that the crossing to negative values of $m_2^2$ shall not occur up to some high scale. In fact, both requirements are not so different, because if we require the $\mathbb{Z}_2$ symmetry to be intact at least to the TeV scale $M_{\rm TeV}= 10^3~{\rm GeV}$ (in order not to disturb DM productions) or to, say, the scale of ``grand unification'' (GUT) $M_{\rm GUT}= 10^{16}~{\rm GeV}$ (where new physics may modify the situation), the corresponding bounds for $\mu_0=M_Z$ are:
\begin{equation}
 m_2(0) \ge \sqrt{\frac{\ln \left(10^3~{\rm GeV}/M_Z, 10^{16}~{\rm GeV}/M_Z \right)}{4 \pi^2}}\,h(0)\,M \approx (0.246, 0.905)\,M\,h(0).
 \label{eq:approx_bound}
\end{equation}
In both cases, as long as the tree level scalar mass parameter $m_2(0)$ is not at least of the size of the heaviest Majorana mass, radiative $\mathbb{Z}_2$ breaking could potentially occur. However, since we at the same time do not want $m_2(0)$ to be too large in order not to pull the Higgs mass parameter to too large values and in the worst case even spoil EWSB, cf.\ Eq.~\eqref{eq:m1RG}, some tension between both requirements is generated.

In the upcoming section we will discuss both the appearance and the impact of radiative $\mathbb{Z}_2$ breaking.

\section{\label{sec:Numerics}Numerical Analysis}

In this section, we will show by a numerical example how non-trivial constraints on the model parameter space can be obtained if the requirement of keeping the $\mathbb{Z}_2$ intact is imposed on some range of energies. We thereby use the full system of RGEs for all three generations. The goal of this section is to illustrate that keeping the $\mathbb{Z}_2$ intact up to a certain energy scale leads to \emph{non-trivial bounds}.

Depending on the setting the $\mathbb{Z}_2$ breaking may be problematic if it, e.g., occurs at scales where DM production could be modified. However, we would like to stress that, due to the size of the parameter space, it is not possible to make any global statements. Rather, whenever a parameter point is studied, one would need to check on a case-by-case basis how/if phenomenology is affected by the resulting constraint.

In general, the model parameters ``beyond the SM'' are $\left\lbrace h_{ij}, M_k, m_2, \lambda_2, \lambda_3, \lambda_4, \lambda_5 \right\rbrace$, some of which are constrained by low- or high-energy data. Nevertheless, we treat these numbers as free parameters, to keep the computational time limited and since e.g.\ a particular leptonic mixing does not affect our considerations. We fix $\lambda_5= 10^{-9}$ at the input scale $\mu = M_Z$, whose value is kept small by the corresponding RGE~\eqref{eq:RGlambda5}. The conditions~\eqref{eq:BoundedFromBelow} require that we use large enough values for $\lambda_2$, such that it is positive for all energy scales. We find that, at $\mu = M_Z$, $\lambda_2 = 0.1$ gives good results while the numerics are under control. For the neutrino Yukawa couplings $h_{ij}$, we input a bimaximally mixed setup~\cite{Barger:1998ta} of $\mathcal{O}(0.1)$ at the GUT scale, which is known to potentially yield phenomenologically valid leptonic mixing at low energies~\cite{Bouchand:2012}.

Moreover, the heavy fields are integrated out for renormalisation scales below their mass thresholds~\cite{Appelquist:1974tg} and the SM input values have been chosen according to~\cite{Agashe:2014kda}.

The connection between the different parameters is intricate: at the input scale, we must choose $m_2^2$ large enough for it not to be driven to \emph{negative} values by the corrections from the Majorana masses, cf.\ Eq.~\eqref{eq:m2RG}. At the same time, a large $m_2^2$ will generate large \emph{positive} corrections to $m_1^2$ [provided that $2\lambda_3 + \lambda_4 > 0$, see Eq.~\eqref{eq:m1RG}], which could spoil EWSB. This is where the actual tension resides: \emph{the running mass parameter $m_2^2$ must lie within the range of the electroweak scale and at the same time it must not be too small to avoid $\mathbb{Z}_2$ symmetry breaking.} There is one more player in the game: the RH neutrino masses drive the tension by their appearance in the RGE for $m_2^2$, Eq.~\eqref{eq:m2RG}. If chosen too large (say, around 10 TeV, for $h_{ij} \sim 0.1$), demanding an unbroken $\mathbb{Z}_2$ symmetry at all scales requires $m_2^2(M_Z)$ to be larger than allowed by the obligation to achieve EWSB.

As mentioned in Sec.~\ref{sec:Vacuum}, the inert doublet VEV may not only lead to breaking of the $\mathbb{Z}_2$, but it could also cause electric charge and/or CP violation. Note however that, since $\lambda_5$ is small for all renormalisation scales, we will have nearly degenerate CP-even and -odd neutral scalar masses, $m_R \simeq m_I$, such that CP violation is inseparable from the breaking of $\mathbb{Z}_2$. To unambiguously identify such scenarios, we have made use of the equivalence of a squared mass becoming negative to symmetry breaking (see Sec.~\ref{sec:Vacuum}). This has been investigated by replacing the Lagrangian parameters in the mass relations~\eqref{eq:masses2} to~\eqref{eq:masses4} by the running couplings and the tree level VEV by a running VEV (cf., e.g., Ref.~\cite{Jegerlehner:2014xxa}):
\begin{equation} \label{eq:running_VEV}
  v^2=-\frac{m_1^2}{\lambda_1} \rightarrow v^2(\mu) = -\frac{m_1^2(\mu)}{\lambda_1(\mu)}.
\end{equation}
This can be done because, in the broken phase, we only need counterterms that are invariant under the broken symmetry group~\cite{collins1984renormalization}. Thus, we can obtain the counterterms in the broken phase from those in the symmetric phase. We can use this fact to construct running quantities in one phase from running quantities in the other~\cite{Kim:2011hx}.

\subsection{\label{sec:Numerics_BScale}Breaking scale}

Let us first try to get a feeling for the numbers involved, i.e., for the scale where the $\mathbb{Z}_2$ breaking occurs as a function of the physical scalar masses.\footnote{We could have used the scalar mass parameters in the Lagrangian, but we consider physical masses to be more illustrative. On top of that, one can use Eqs.~\eqref{eq:masses2} to~\eqref{eq:masses4} to convert one into the other.}

\begin{figure}[t]
\hspace{-1cm}
\begin{tabular}{lr}
  \includegraphics[trim=1cm 1cm 1cm 1cm,width=8cm]{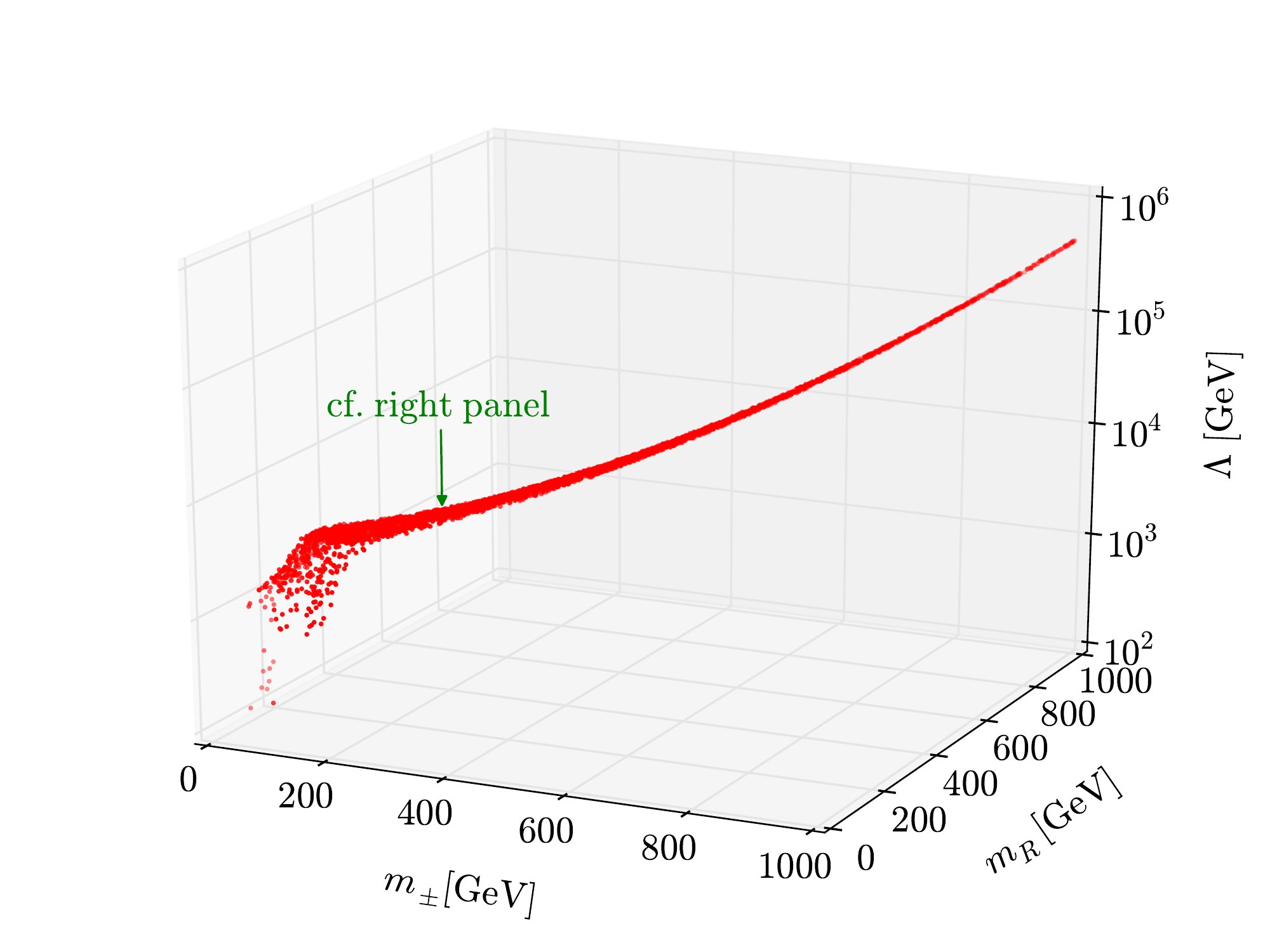} & \includegraphics[trim=0cm 1cm 1cm 1cm, width=8cm]{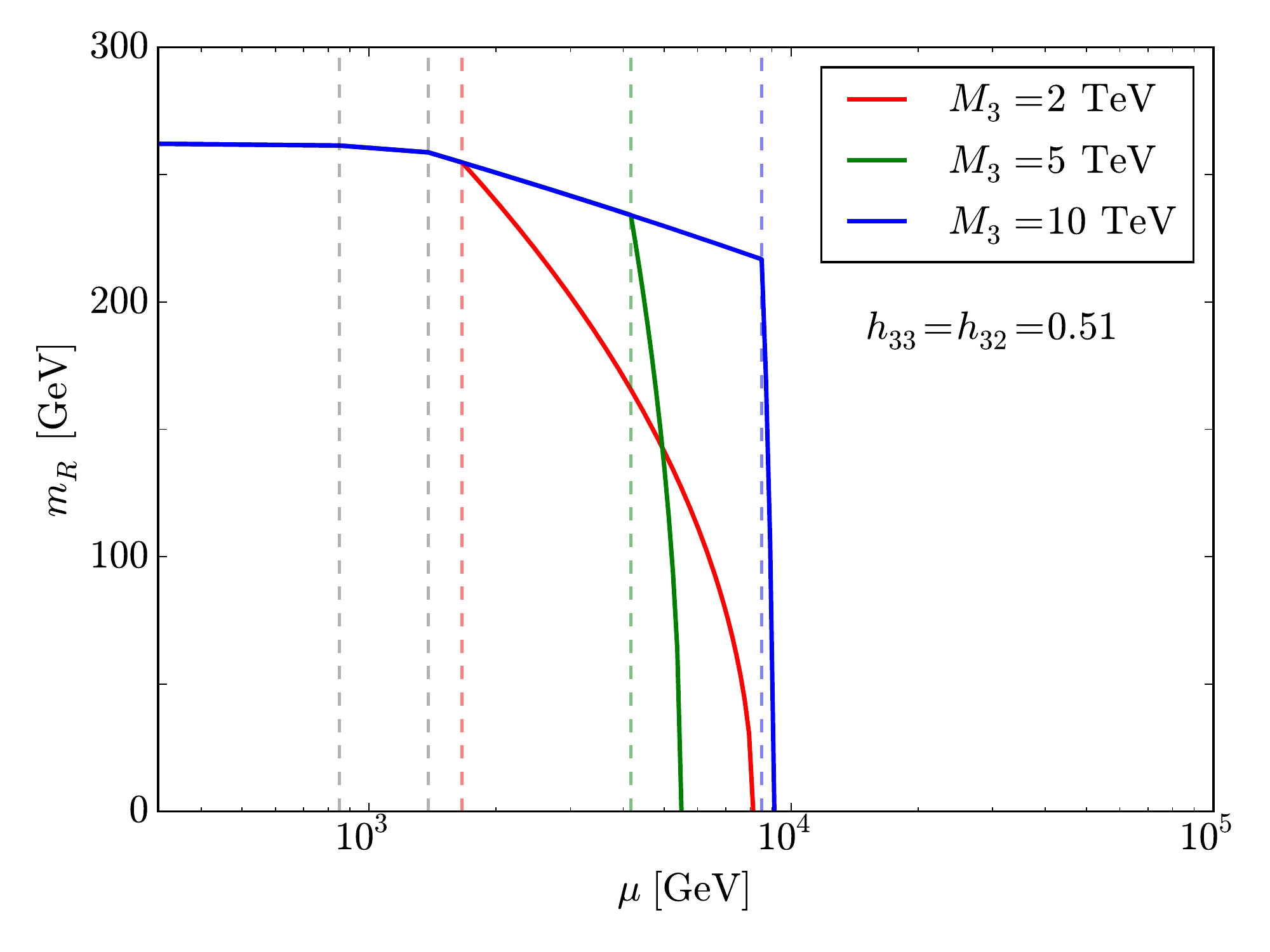}
  \end{tabular}
  \caption{\label{fig:breaking}\emph{Left}: $\mathbb{Z}_2$-breaking scale $\Lambda$ as a function of the masses $m_{\pm,R}$ for RH masses $\left.(M_1, M_2, M_3)\right|_{\mu=M_\textrm{GUT}} = (900, 1500, 5000)$~GeV. 
  \emph{Right}: Explicit running of \mbox{$m_R\ (<m_\pm)$} for one example set of parameters with $\left.(M_1,M_2)\right|_{\mu=M_\textrm{GUT}} = (900, 1500)$~GeV (dashed grey lines). In both plots, Yukawa couplings $<1$ are chosen (the crucial ones are given).}
\end{figure}

The natural question is at which scale the breaking could occur for realistic particle masses. The crucial observation is that even RH neutrino masses in the TeV range could be perfectly sufficient to break the $\mathbb{Z}_2$ parity at a relatively low scale, as Eqs.~\eqref{eq:condition1} and~\eqref{eq:condition2} suggest. To illustrate this statement, we present in the left panel of Fig.~\ref{fig:breaking} the breaking scale as a function of the scalar masses $m_\pm$ and $m_R$. We have neglected all SM fermions but the top quark to be able to determine the breaking scale with higher precision. The RH neutrino masses are $(M_1, M_2, M_3)=(900, 1500, 5000)$~GeV, input at $M_\textrm{GUT}$, with Yukawa couplings of $\mathcal{O}(0.1)$, which are all not overly large and more or less in the ranges where many particle physicists would ``naturally'' expect them. However, glancing at the figure, it is visible that even for these values the breaking scale can be as low as a few TeV. The breaking scale grows with both 
$m_\pm$ and $m_R$, as to be expected from Eqs.~\eqref{eq:masses2} and~\eqref{eq:masses3}, since both of them grow with $m_2$ but are related as $m_R^2 - m_\pm^2 = (\lambda_4 + \lambda_5) v^2$. Ultimately, the value of the breaking scale is determined by a tug-of-war between the scalar mass parameter $m_2$ and the maximum of all $\left\lbrace M_i h_{ij}\right\rbrace$, cf.\ Eq.~\eqref{eq:approx_bound}.

To see the exact influence of the combination of Yukawa coupling and mass, we can look at the right panel of Fig.~\ref{fig:breaking}, where the explicit evolution of $m_R$ is shown as a function of the renormalisation scale, for three different exemplary masses $M_3$. Indeed, as soon as $M_3$ becomes dynamic, it pulls $m_R$ towards smaller values, until it is zero. Beyond this scale $m_R^2$ is negative, i.e.\ the ``physical'' mass $m_R$ becomes imaginary. However, given that this occurs in a setting with $m_1^2<0$ and $m_R < m_\pm$, the scale at which $m_R=0$ is to be identified with the $\mathbb{Z}_2$ breaking scale, so that in fact the assumed field configuration is no longer expanded around a minimum of the potential for larger renormalisation scales. Expanding around an actual minimum would of course yield real and positive masses.

An important point to make at the end this section is to understand that it is not immediately visible whether or not the breaking emerges at sufficiently low scales and destroys the validity of the model. Instead, one has to look at the details of the parameters under consideration, and if in doubt apply further analyses that allow to clarify the situation, as we shall outline in Sec.~\ref{sec:Discussion}. However, as can be seen from the right panel of Fig.~\ref{fig:breaking}, it is well possible that e.g.\ for an inert scalar with a mass of 262~GeV, which is the DM candidate of that setting if it is the lightest electrically neutral and stable particle involved, the symmetry breaking can already occur at about 1~TeV, which is very close to the phase decisive for DM production in the early Universe. In such a case, the running of the parameters involved cannot simply be neglected when computing the DM production. This observation is also true if the effect of $\mathbb{Z}_2$ breaking is 
related to thermal effects, which may significantly alter the situation \cite{Land:1992sm, Gil:2012ya, Patel:2012pi, Blinov:2015sna, Blinov:2015vma}, as we will explicitly discuss in Sec.~\ref{sec:Discussion}.

\subsection{\label{sec:Numerics_WayOut}A humble way out}

 For some studies it may not be easy to show whether or not a too low breaking scale has a negative impact, e.g., if no suitable tools to compute DM production are available.\footnote{While DM production can be handled easily in simple cases, in particular settings with co-annihilations or non-thermal production are more involved.} In other contexts one may only be interested in a rough picture of the situation, where the decisive point is to avoid a potential symmetry breaking altogether. In such cases, a convenient way out would be to require the breaking scale to be at such high energies that it does not affect low-energy physics. For example, one could ask for which parameter values the $\mathbb{Z}_2$ breaking happens only beyond $M_{\rm GUT}$, where new physics would potentially modify the situation. In the following, we give an example for how this would affect the parameter space available.
 
 To illustrate the parity problem, we generated a random set of $10^5$ input values within:
 \begin{equation}
  0\,\textnormal{TeV} \le m_2(M_Z) \le 3\,\textnormal{TeV}, \quad -1 \le \lambda_3(M_Z), \lambda_4(M_Z) \le 1,
 \end{equation}
 for masses $( M_1,M_2,M_3 ) = ( 900,1500,2000)\, \textnormal{GeV}$ as input at $M_\textrm{GUT}$. We then computed the running effects for each point chosen, this time taking into account all SM fermions. First, we solve the RGEs for the running couplings of the scalar sector ($\lambda_{1,2,3,4,5}$) and check the scalar consistency criteria, Eq.~\eqref{eq:BoundedFromBelow}, for all energy scales between $M_Z$ and $M_{\rm GUT}$.\footnote{This ensures that we do not have to consider further particle thresholds and all SM fermions can be safely neglected. This even holds for the top quark given that its mass is so close to the lower input scale.} If these are not violated for any scale below $M_{\rm GUT}$, we solve the scalar mass parameter RGEs~\eqref{eq:m1RG} and~\eqref{eq:m2RG}. If we do not find EWSB below $1$~TeV, i.e.\ if $m_1^2(\mu)\ge 0$, we reject the input values. Otherwise we distinguish two sub-cases depending on the sign of $m_1^2(\mu)$, see Eq.~\eqref{eq:Z2broken}. In case we find that all vacua 
that respect the $\mathbb{Z}_2$ parity are unstable, we also reject the input values. By adhering to this approach, we can avoid rejecting input values because of a broken $\mathbb{Z}_2$ symmetry although some other criterion fails as well, i.e., we determine exactly those points which would \emph{not} be rejected if one disregarded the bound arising from radiative symmetry breaking.

\begin{figure}[t]
\centering
\begin{tabular}{lr}
  \includegraphics[width=8cm, trim=25mm 0 25mm 0]{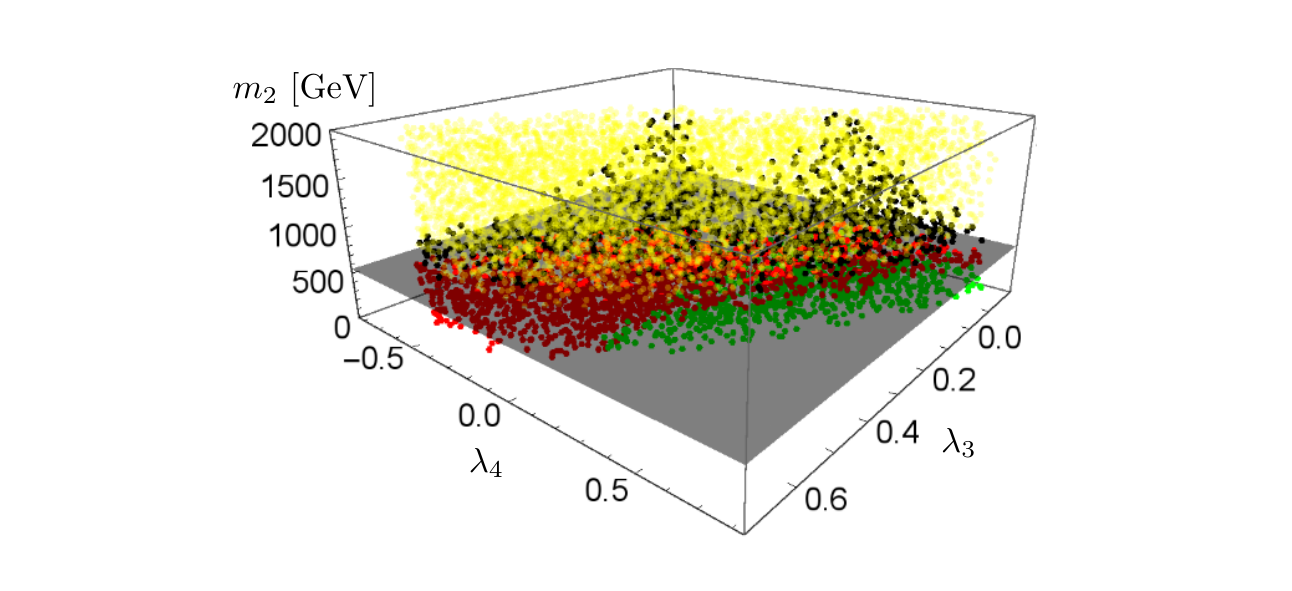}
&
  \raisebox{3mm}{
  \includegraphics[width=8cm, trim=25mm 0 25mm 0]{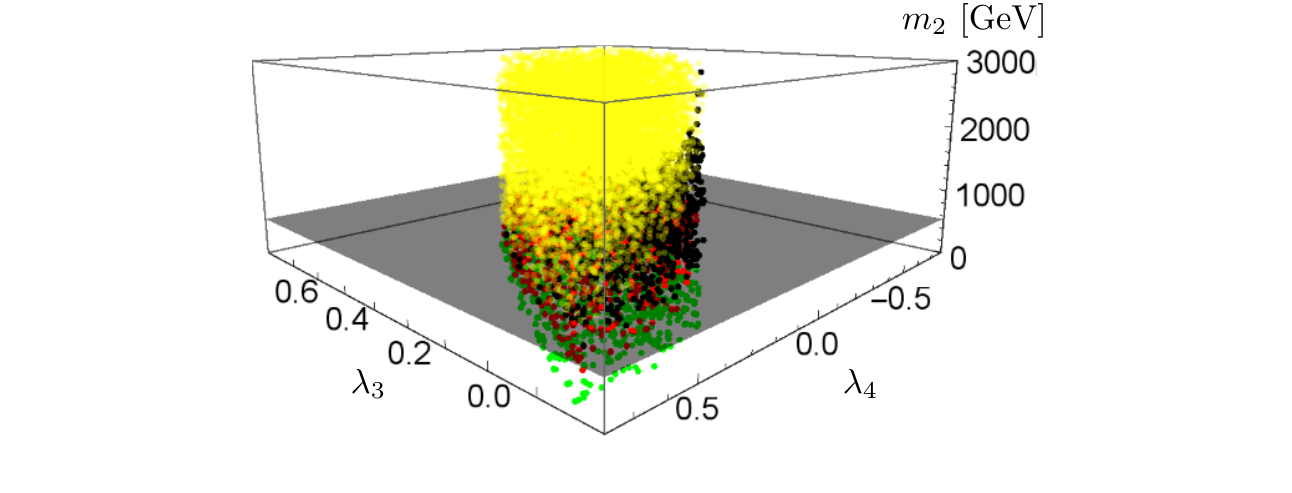}}
\end{tabular}
  \caption{\label{fig:result1}Parameter scan for the scotogenic model. Black dots are valid input parameter values, yellow dots violate EWSB, red ($m_{2/R}^2<0$) and green ($m_\pm^2<0$) dots are excluded due to breaking of $\mathbb{Z}_2$. The shaded grey area represents the ``lower bound'' on $m_2$.}
\end{figure}

The results of our parameter scan are shown in Fig.~\ref{fig:result1}, where the allowed and rejected input values at the electroweak scale are marked.\footnote{This time, we have chosen to display the Lagrangian parameters as opposed to Fig.~\ref{fig:breaking}, since the resulting bound is illustrated in a better way.} All data points violating vacuum stability or perturbativity have been excluded from the plot. The colour code is as follows: \emph{Black dots} mark the points which fulfil all the constraints. All other points are excluded for various reasons. \emph{Yellow dots} are excluded due to failing to produce EWSB below $1$~TeV, while \emph{red/green dots} are excluded \emph{only} because they lead to radiative $\mathbb{Z}_2$ breaking, where $m_{2/R}^2 < 0$ is signified by red dots and $m_\pm^2 < 0$ by green ones.

For a large range of input values for $\lambda_3$ and  $\lambda_4$, the criterion that we must encounter EWSB below 1~TeV translates into an upper bound on $m_2(M_Z)$, which is a function of $\lambda_3$ and $\lambda_4$. Only if $2\lambda_3 + \lambda_4 < 0$, EWSB can be maintained below 1~TeV while allowing for a large $m_2$. This manifests itself in the columns of valid (black) data points ranging up to very high input values of $m_2$, which are visible in both panels of Fig.~\ref{fig:result1}.

On the other hand, one can see that including the criterion of an unbroken $\mathbb{Z}_2$ symmetry is essentially equivalent to a lower bound on $m_2(M_Z)$ -- in our example about $550\,\textnormal{GeV}$, indicated by the grey area in the plots. As can be seen in the left Fig.~\ref{fig:result1}, we have found red and green points \emph{below} this bound, which are only excluded by the requirement of the $\mathbb{Z}_2$ symmetry to be unbroken at all scales. Even above this bound it is visible that, for certain choices of $\lambda_{3,4}$, there are some red and green dots which would be considered unproblematic if parity breaking was disregarded.

For the physical scalar masses in the minimal subtraction (MS) scheme, we obtain values in the ranges 
\begin{eqnarray}
 &&554.2\textrm{ GeV} \le m_\pm(\mu=m_\pm) \le 2780.6\textrm{ GeV}, \nonumber\\
 &&558.5\textrm{ GeV} \le m_{R/I}(\mu=m_{R/I}) \le 2781.6\textrm{ GeV},\label{eq:scen_a_ranges}
\end{eqnarray}
confirming our expectation of a lower bound on the masses. Generalising our estimate from Eq.~\eqref{eq:approx_bound} to the case of three generations, we find agreement with the exact result:
\begin{equation}
 0.905\ \sqrt{\max_i \left|M_i(M_Z)^2 \left[h(M_Z) h^\dag(M_Z)\right]_{ii}\right|} \approx 625\, {\rm GeV},
\end{equation}
to be compared to the estimated value of $550$~GeV. At first sight it may be surprising that the simple estimate outlined in Sec.~\ref{sec:Vacuum} leads to such a good agreement. However, given that there is quite a range possible for both Yukawa couplings and RH neutrino masses, it is in fact to be expected that, rather generically, one of the products of the form (neutrino Yukawa coupling) $\times$ (RH neutrino mass) will in most cases dominate over the others, thereby mimicking the situation of only one RH neutrino being present.

Glancing at Eq.~\eqref{eq:m2RG}, one might be led to the assumption that, by raising the Majorana masses, the transition to negative values of $m_2^2$ can be pushed beyond $M_{\rm GUT}$, since below their mass thresholds the RH neutrino fields are integrated out. However such an attempt must fail since it is always overcompensated by the quadratic term in the RGE for $m_2^2$, Eq.~\eqref{eq:m2RG}, which is also illustrated by Fig.~\ref{fig:breaking}, right panel. Only if we chose all $M_i \ge M_{\rm GUT}$ could it be achieved, but at latest at that point some other new physics would probably appear which may completely change the situation. 

Having seen the examples in this section, it is evident that potentially non-trivial constraints can arise from avoiding the $\mathbb{Z}_2$ breaking. However, depending on the context they may be stronger or weaker, so that the ultimate conclusion is that, for any given parameter point leading to $\mathbb{Z}_2$ breaking, one must check whether or not the breaking may have a bad influence, in particular on the production of DM.

\section{\label{sec:Discussion}Discussion \& Outlook}

In this section, we will briefly discuss some further phenomenological implications and/or subtleties related to the parity problem. As we will see, none of the points changes the principal situation, however, the points discussed may lead to further constraints or they may be worth to be considered in a separate work.

\subsection{Unstable vacuum}

As we stated in Sec.~\ref{sec:Vacuum}, one can in principle obtain stronger constraints on the scalar sector if demanding that one of the $\mathbb{Z}_2$ symmetric minima is also the global one. If this were not the case, the local minima would simply decay into the global vacuum. 

\begin{figure}[t]
  \centering
  \includegraphics[width=.66\textwidth]{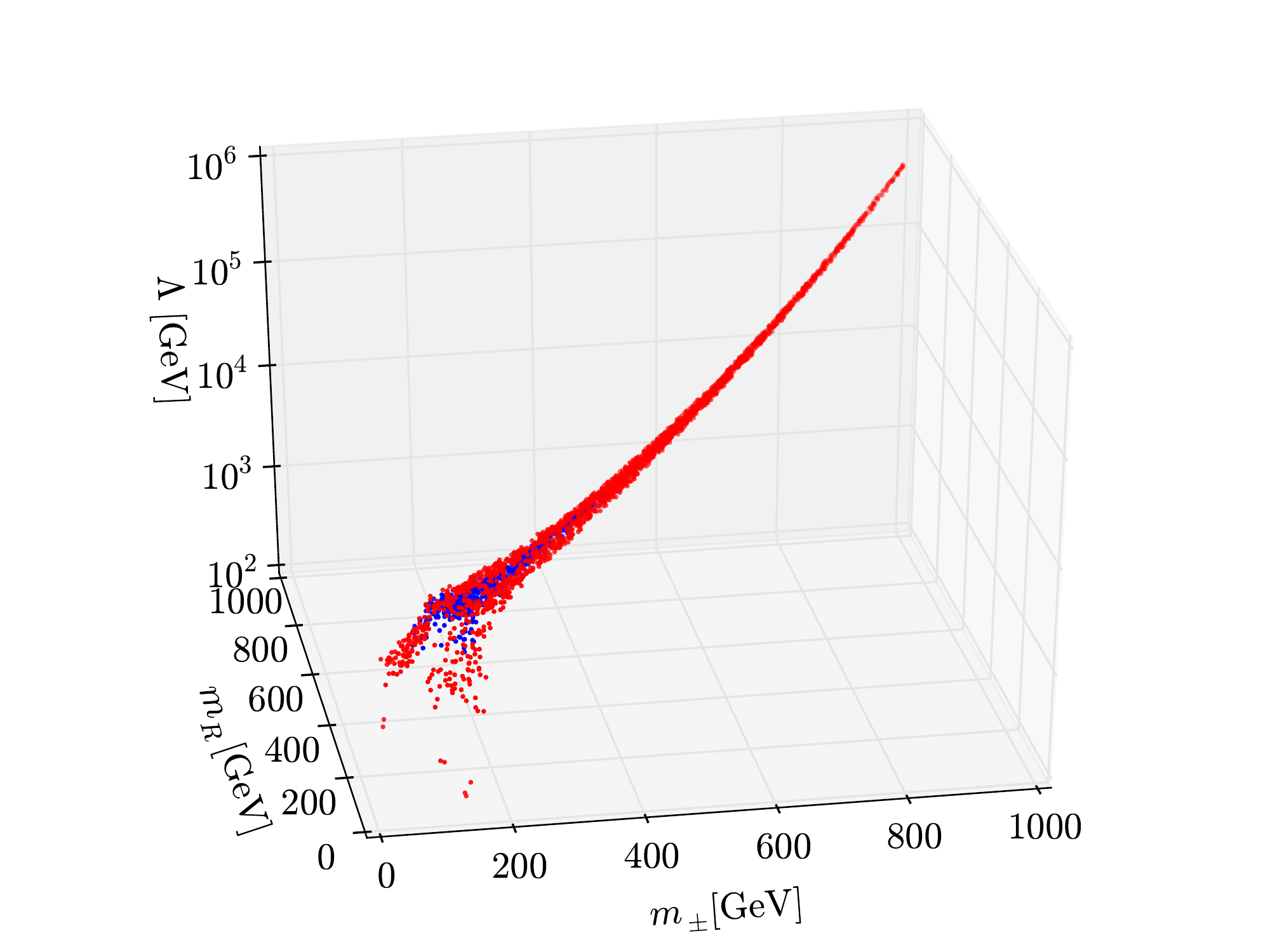}
  \caption{\label{fig:Z2BreakingGlobal}Re-run of the scan shown in Fig.~\ref{fig:breaking} (left panel), where \emph{blue} points indicate that the $\mathbb{Z}_2$ symmetric vacua are \emph{not} the global ones.}
\end{figure}

In repeating the analysis presented in Fig.~\ref{fig:breaking} (left panel) and checking for this additional feature, we obtain Fig.~\ref{fig:Z2BreakingGlobal}. Here, the red points indicate that $\mathbb{Z}_2$ breaking is the only constraint, while the blue points indicate that there is a deeper minimum emerging at a scale \emph{below} the breaking scale $\Lambda$, thus giving even stronger constraints. As we can see, this is only relevant for light scalar masses. For large mass scales, $\mathbb{Z}_2$ breaking evidently occurs before deeper minima can form. 

Unfortunately, it is rather difficult to infer any global qualitative statements from Eqs.~\eqref{eq:GlobalMin1} to~\eqref{eq:GlobalMin4} since they depend on the concrete values of the scalar couplings $\lambda_{1,2,3,4,5}$ at a given scale. Intuitively, one would expect new minima to occur if $m_2^2$ is either negative or at least small. For large physical scalar masses, $m_2^2$ will be positive and large. Thus, given that the scalar couplings have a tendency to run to smaller values for high scales, the scale of $\mathbb{Z}_2$ breaking for growing $m_2^2$ approaches the scale where new minima emerge.\footnote{For example, the Higgs quartic coupling $\lambda_1$ is dominated by the top Yukawa coupling $y_t$ which gives a negative contribution, cf.\ Eq.~\eqref{eq:RGlambda1}.} Note that by a \emph{small} mass parameter in this context we refer to the different combinations of mass parameters and quartic couplings, e.g.\ $\frac{m_{1}^4}{\lambda_{1}}$ in Eq.~\eqref{eq:GlobalMin2}.

\subsection{DM decay}

We have already mentioned that DM production may be affected by the parity problem. It is worth to note that the issue is however not so much about the mere abundance of the DM. While indeed a broken $\mathbb{Z}_2$ symmetry would change the Feynman rules in the scotogenic model, and hence the amount of DM produced in the early Universe, this could typically be compensated by choosing slightly different parameter values which, given that neither the inert scalar mass nor the RH neutrino masses are known accurately, is not expected to be a major issue.

The actual problem lies somewhere else: given that the radiatively induced VEV $v_\eta$ of $\eta$ is not tiny (i.e., it can be expected to be at least in the GeV to TeV ballpark for generic parameter choices) and that the coupling strengths are not extremely small, the resulting DM candidate will generically decay very quickly. While decaying DM is not excluded by observations, its interactions typically need to be strongly suppressed e.g.\ by very heavy mediators or by tiny ($\sim$ gravitational) interaction strengths -- see Ref.~\cite{Ibarra:2013cra} for a discussion of some example models. If this is not the case, the DM is in great danger to simply ``decay away'', i.e., to have a lifetime smaller than the age of the Universe, so that it could not have survived until today when produced early in history. Potentially fast decays would furthermore tend to increase the freeze-out temperature, since decaying DM drops out of thermal equilibrium earlier than stable DM, so that on top of that (at least for non-
relativistic DM particles) a strong modification of the abundance can be 
expected, too.

One can easily see that the constraint arising from the lifetime is rather strong. For example, taking $v_\eta \lambda$ as Feynman rule for the vertex which allows an inert scalar to decay into two SM-like Higgses, the resulting decay width could be approximately computed to be $\Gamma \sim v_\eta^2 \lambda^2 / (16 \pi m_\eta)$. If we conservatively estimate the DM lifetime to be at least larger than the age of the Universe ($13.81$~Gyr~\cite{Agashe:2014kda}), corresponding to about $4.3\cdot 10^{17}$~sec, we get an upper limit of about $1.5\cdot 10^{-27}$~GeV on the decay rate. Even for a mass of $m_\eta = 500$~GeV and a ``small'' VEV of $v_\eta = 1$~GeV, the upper limit on the interactions strength would then be $\lambda \lesssim 6\cdot 10^{-12}$. For a scalar potential like the one in Eq.~\eqref{eq:potential}, this bound would extend to all three couplings $\lambda_{3,4,5}$, at least in the absence of very fine-tuned cancellations. Taking into account some information on the decay products, this limit 
would become even stronger~\cite{Bell:2010fk}. Similar estimates could be done in case one of the RH neutrinos was the DM particle.

Hence, unless we resort to extremely small couplings or very strong fine tuning, a VEV of the inert scalar would quite generally destroy the ability of the scotogenic model to explain DM, simply because the interactions involved cannot be sufficiently suppressed.

\subsection{Thermal effects}

\begin{figure}[t]
  \centering
  \includegraphics[width=.66\textwidth]{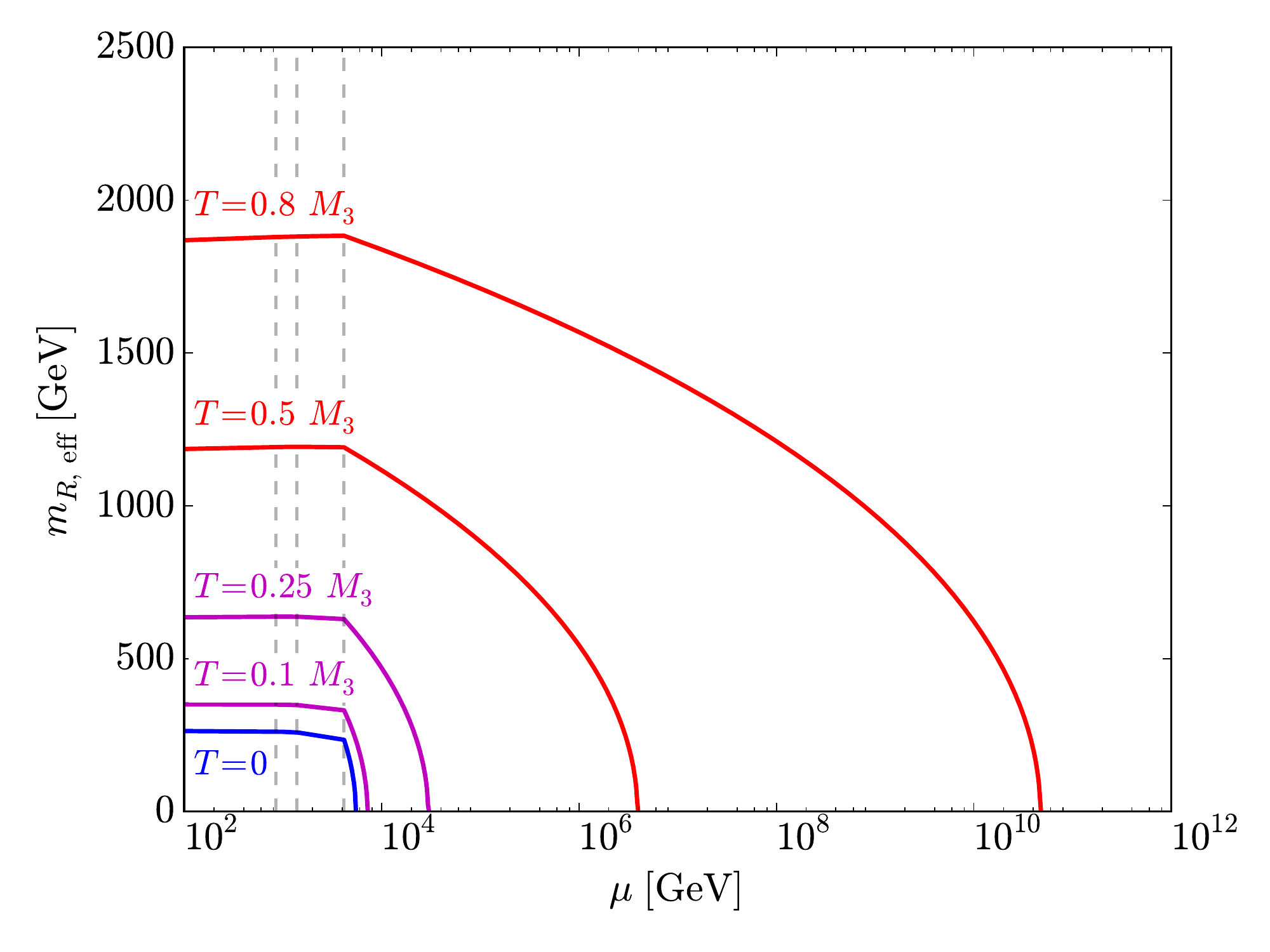}
  \caption{\label{fig:Z2BreakingTemp}Effective thermal mass parameter $m_{R,\, \textrm{eff}}$ for different temperatures $T$.}
\end{figure}

Thermal effects can be important for our analysis since broken symmetries will generally be restored at sufficiently high temperatures~\cite{Quiros:1999jp}. One might be led to the conclusion that this makes the entire discussion of $\mathbb{Z}_2$ breaking superfluous, given that the breaking occurs at high scales only. This is, of course, too simple an argument since the real situation is much more complex. To understand this, let us reconsider the situation shown in Fig.~\ref{fig:breaking} (right panel) with a $5\textrm{ TeV}$ heaviest RH neutrino and $\eta_R$ the lightest inert particle. We can capture the leading thermal effects by including the thermal corrections to the quadratic terms in the scalar potential~\cite{Ginzburg:2010wa}. This is achieved by a shift in the mass parameter $m_2^2 \to m_2^2 + c T^2$, where $T$ is the temperature and c is given by~\cite{Gil:2012ya, Blinov:2015vma}:
\begin{equation}
  c = \frac{1}{16} \left(g_1^2 + 3 g_2^2\right) + \frac{\lambda_2}{4} + \frac{1}{12} \left( 2\lambda_3 + \lambda_4\right) + \frac{1}{12} \mathrm{tr}\left(h^\dag h\right),
\end{equation}
as can be calculated from the IR limit of the thermal scalar two-point function~\cite{Carrington:1991hz,Parwani:1991gq,Arnold:1992rz}. In this way, we obtain the effective thermal mass parameter $m_{R,\,\textrm{eff}}$, which is shown in Fig.~\ref{fig:Z2BreakingTemp} for different temperatures. Clearly, at high temperatures the $\mathbb{Z}_2$ symmetry will be restored for a fixed renormalisation scale $\mu$, as one would expect. However, since our DM candidate in this scenario is $\eta_R$, which is a weakly interacting massive particle (WIMP) with a physical mass around $250\textrm{ GeV}$, the temperatures relevant for the thermal decoupling will be much \emph{below} $M_3$, such that the thermal corrections have only little impact (cf.\ violet curves). The remaining question of how to choose the renormalisation scale can only be answered if one looks at the details of the DM production. We leave such details to be worked out in future studies.

In combination with the discussion on DM decay, this shows that thermal effects will alter the details but not the overall picture: $\mathbb{Z}_2$ breaking can occur and may have a significant impact on the phenomenology of the scotogenic model. Any phenomenological study should include this fact in their considerations.

\section{\label{sec:Summary}Conclusions}

We have illustrated that the scotogenic neutrino mass model suffers from a \emph{parity problem}, i.e., it is in danger that its intrinsic built-in $\mathbb{Z}_2$ parity symmetry is broken by quantum effects driven by the heavy right-handed neutrinos. This could generate unwanted effects such as modifications of DM production. This issue imposes visible constraints on the parameter space available, in particular because the most generic solution, i.e., simply pushing the corresponding mass parameter in the Lagrangian to large enough values to avoid the breaking, does \emph{not} work due to electroweak symmetry breaking being threatened. Thus, the scotogenic model suffers from tension from two different sides which considerably shrinks the allowed parameter ranges.

After introducing the scotogenic model and its vacuum structure, we have presented an analytical approximation to compute the breaking scale, before exemplifying the resulting constraints numerically. Our considerations are based on the 1-loop renormalisation group equations of the scotogenic model which we have re-derived and, in passing, updated compared to previous versions in the literature. Summing up, we have revealed a somewhat subtle but non-trivial constraint on the scotogenic model which is able to strongly reduce the allowed parameter space. This makes it necessary for future considerations to check whether the parity problem exists for a certain choice of parameters, or not, to avoid the trap of studying physically irrelevant regions of the model.

\section*{Acknowledgements}

We are indebted to Patrick Otto Ludl, Michal Malinsky, and Jose Miguel No for very enlightening discussions, and we are particularly grateful to Nicolas Rojas, Jose W.~F.~Valle, and Avelino Vicente for useful comments on the manuscript, an extended discussion of the implications of discrete symmetry breaking in the scotogenic model, and for cross-checking our RGEs. AM acknowledges partial support by the European Union FP7 ITN-INVISIBLES (Marie Curie Actions, PITN-GA-2011-289442).

\appendix

\section{\label{app:Appendix_RGE}Renormalisation group equations}
\renewcommand{\theequation}{A-\arabic{equation}}
\setcounter{equation}{0}  

The 1-loop RGEs for the scotogenic model have first been computed in Ref.~\cite{Bouchand:2012}. We have re-derived those equations needed for the purpose of this paper, and have in passing taken the opportunity to update part of the earlier results.

For convenience, we define the differential operator $\mathcal{D} \equiv 16 \pi^2 \mu \frac{\textrm{d}}{\textrm{d}\mu}$. The 1-loop RGEs for the gauge couplings are those of a generic THDM~\cite{Grzadkowski:1987wr}:
\begin{equation}
\mathcal{D} g_i = b_i g_i^3 \textnormal{ (no sum!)},
\label{eq:gauge-RGE}
\end{equation}
with $b = \left( 7,-3,-7 \right)$.

The quark sector of the scotogenic model is the same as that of the SM, such that the corresponding RGEs do not change.\footnote{Note the implicit changes in $g_{1,2}$, though, by virtue of Eq.~\eqref{eq:gauge-RGE}.} The RGEs for the leptonic Yukawa couplings are:
\begin{subequations}
\begin{align}
 \mathcal{D} Y_e &= Y_e \left\lbrace \frac{3}{2}Y_e^\dagger Y_e +\frac{1}{2} h^\dagger h + T - \frac{15}{4} g_1^2 - \frac{9}{4} g_2^2 \right\rbrace, \label{eq:leptonRG}\\
 \mathcal{D} h &= h \left\lbrace \frac{3}{2} h^\dagger h + \frac{1}{2} Y_e^\dagger Y_e + T_\nu - \frac{3}{4} g_1^2 - \frac{9}{4} g_2^2 \right\rbrace,
 \label{eq:neutrinoRG}
\end{align}
\end{subequations}
where $T_\nu \equiv \textrm{Tr}\left(h^\dagger h \right)$ and $T \equiv \textrm{Tr}\left(Y_e^\dagger Y_e + 3 Y_u^\dagger Y_u + 3 Y_d^\dagger Y_d\right)$. 
For the Majorana mass matrix, one finds~\cite{Antusch:2002rr,Bouchand:2012}:
\begin{equation}
 \mathcal{D} M = \left\lbrace \left(h\, h^\dagger\right) M + M \left(h\, h^\dagger \right)^* \right\rbrace.
 \label{eq:RHnuRG}
\end{equation}
For the quartic scalar couplings, we find the RGEs for a $\mathbb{Z}_2$ symmetric THDM~\cite{Hill:1985}:
\begingroup
\setlength{\jot}{8pt}
\begin{subequations}
\begin{align} 
 \begin{split} \label{eq:RGlambda1}
     \mathcal{D} \lambda_1 &= 12 \lambda_1^2 + 4 \lambda_3^2 + 4 \lambda_3 \lambda_4 + 2 \lambda_4^2 + 2 \lambda_5^2 
	+ \frac{3}{4} \left(g_1^4 + 2 g_1^2 g_2^2 + 3 g_2^4\right) \\
	&\qquad - 3 \lambda_1 \left(g_1^2 + 3g_2^2\right) + 4 \lambda_1 T - 4 T_4,
 \end{split}\\
 \begin{split} \label{eq:RGlambda2}
  \mathcal{D} \lambda_2 &= 12 \lambda_2^2 +4 \lambda_3^2 + 4 \lambda_3 \lambda_4 + 2 \lambda_4^2 + 2 \lambda_5^2 
	+ \frac{3}{4} \left(g_1^4 + 2 g_1^2 g_2^2 + 3 g_2^4\right) \\
	&\qquad - 3 \lambda_2 \left(g_1^2 + 3g_2^2\right) + 4 \lambda_2 T_\nu - 4 T_{4\nu},
 \end{split} \\
 \begin{split} \label{eq:RGlambda3}
  \mathcal{D} \lambda_3 &= 2 \left(\lambda_1 + \lambda_2 \right) \left( 3\lambda_3 + \lambda_4\right) + 4\lambda_3^2 + 2 \lambda_4^2 + 2 \lambda_5^2 
	+ \frac{3}{4} \left(g_1^4 - 2 g_1^2 g_2^2 + 3 g_2^4\right) \\
	&\qquad - 3 \lambda_3 \left(g_1^2 + 3g_2^2\right) + 2 \lambda_3 \left(T +T_\nu\right) - 4 T_{\nu e},
 \end{split} \\
 \begin{split} \label{eq:RGlambda4}
  \mathcal{D} \lambda_4 &= 2 \left(\lambda_1 + \lambda_2 \right) \lambda_4 + 8 \lambda_3 \lambda_4 + 4 \lambda_4^2 + 8 \lambda_5^2 + 3 g_1^2 g_2^2 \\
	&\qquad - 3 \lambda_4 \left(g_1^2 + 3g_2^2\right) + 2 \lambda_4 \left(T +T_\nu\right) + 4 T_{\nu e},
 \end{split} \\
  \mathcal{D}\lambda_5 &= \lambda_5 [ 2\left(\lambda_1 + \lambda_2\right) + 8\lambda_3 +12\lambda_4 
	- 3 \left(g_1^2 + 3g_2^2\right) + 2 \left(T + T_\nu\right)],  \label{eq:RGlambda5}
\end{align}
\end{subequations}
\endgroup
where we have used the abbreviations $T_{4}\equiv \mathrm{Tr} \left( Y_e^\dag Y_e Y_e^\dag Y_e + 3 Y_u^\dag Y_u Y_u^\dag Y_u + 3 Y_d^\dag Y_d Y_d^\dag Y_d \right)$, $T_{4\nu}\equiv \mathrm{Tr} \left( h^\dag h \, h^\dag h \right)$ and $T_{\nu e}\equiv \mathrm{Tr} \left( h^\dag h \,Y_e^\dag Y_e \right)$.

\noindent
The scalar mass parameters obey the following RGEs:
\begin{subequations}
\begin{align} \label{eq:m1RG}
  \mathcal{D}m_1^2 &= 6 \lambda_1 m_1^2 +2\left(2\lambda_3 + \lambda_4\right)m_2^2 + m_1^2\left[ 2T - \frac{3}{2} \left(g_1^2 + 3g_2^2\right)\right],\\
  \label{eq:m2RG}
    \mathcal{D}m_2^2 &= 6 \lambda_2 m_2^2 +2\left(2\lambda_3 + \lambda_4\right)m_1^2  
	+ m_2^2\left[ 2T_\nu - \frac{3}{2} \left(g_1^2 + 3g_2^2\right)\right] - 4 \sum_{i=1}^3 M_i^2\left(h \, h^\dagger\right)_{ii},
\end{align}
\end{subequations}
where the last term in Eq.~\eqref{eq:m2RG} is characteristic for a scalar field coupled to Majorana fermions (see, e.g., Ref.~\cite{Kersten:2001,Vissani:1997ys}). This term is the crucial point of our study. Conveniently, since it is nothing but a trace, it is invariant under the transformation that diagonalises $M$, such that we do not have to perform this diagonalisation explicitly. Furthermore, the decoupling of the heavy Majorana fields must be carried out by hand, since we are working in a mass-independent renormalisation scheme. To this end, we match the neutrino Yukawa couplings and Majorana mass matrices in the basis where $M$ is diagonal \cite{Antusch:2002rr}. Apart from these two quantities, the only appearance of the Majorana masses is Eq.~\eqref{eq:m2RG}, where it suffices to remove the corresponding contribution.

\bibliographystyle{./apsrev}
\bibliography{./literature}

\end{document}